\documentclass[prl,10pt,a4paper,superscriptaddress,twocolumn,showpacs,noshowkeys]{revtex4}

\usepackage{times}
\usepackage[dvipdfm]{graphicx}
\usepackage{amsmath}
\usepackage{amssymb}

\begin{document}

\title{Nonlinear Schr\"{o}dinger equation with chaotic, random, and nonperiodic nonlinearity}
\author{W. B. Cardoso}
\affiliation{Instituto de F\'{\i}sica, Universidade Federal de Goi\'{a}s, 74001-970, Goi%
\^{a}nia, GO, Brazil.}
\author{S. A. Le\~{a}o}
\affiliation{Instituto de F\'{\i}sica, Universidade Federal de Goi\'{a}s, 74001-970, Goi%
\^{a}nia, GO, Brazil.}
\author{A. T. Avelar}
\affiliation{Instituto de F\'{\i}sica, Universidade Federal de Goi\'{a}s, 74001-970, Goi%
\^{a}nia, GO, Brazil.}
\author{D. Bazeia}
\affiliation{Departamento de F\'{\i}sica, Universidade Federal da Para\'{\i}ba,
58051-970, Jo\~{a}o Pessoa, PB, Brazil.}
\author{M. S. Hussein}
\affiliation{Departamento de F\'{\i}sica Matem\'{a}tica, Instituto de F\'{\i}sica, Universidade
de S\~{a}o Paulo, 05314-970, S\~{a}o Paulo, SP, Brazil} 


\begin{abstract}
In this paper we deal with a nonlinear Schr\"{o}dinger equation
with chaotic, random, and nonperiodic cubic nonlinearity. Our goal is
to study the soliton evolution, with the strength of the nonlinearity perturbed
in the space and time coordinates and to check its robustness under
these conditions. Comparing with a real system, the perturbation can
be related to, e.g., impurities in crystalline structures, or coupling
to a thermal reservoir which, on the average, enhances the nonlinearity.
We also discuss the relevance of such random perturbations to the dynamics
of Bose-Eisntein Condensates and their collective excitations and transport.
\end{abstract}

\maketitle

The nonlinear Schr\"{o}dinger equation (NLSE) is the mathematical vehicle
that describes the evolution of solitonic solutions for different nonlinear systems,
such as, fiber optics \cite{Agrawal01}, bulk medium and photonic crystals \cite{Kivshar03},
Langmuir waves in plasmas \cite{ZakharovJEPT72}, wave function of
Bose-Einstein condensates (BECs) \cite{Pethick02}, and others.

A special case involving the NLSE consists in variable coefficients
modulated in the spatial and/or temporal coordinates. The control of
these coefficients allow us to obtain new distinct solutions. 
In this context, \cite{Belmonte-BeitiaPRL08,AvelarPRE09} have recently
proposed a treatment of BECs using similarity transformations to
construct explicit nontrivial solutions of the cubic and cubic-quintic NLSE
with potentials and nonlinearities depending both on time and on the spatial
coordinates. Also, thermal effects
on nonlinearities can change its form, presenting new solutions \cite%
{GharaatiAPPA07}. In BECs, the $s$-wave scattering length of interatomic
collisions determines the strength of the nonlinearity coefficient \cite{Pethick02}, and it
can be controlled using the Feshbach resonance (FR) \cite{TTHK99} via external magnetic \cite%
{KohlerRMP06,InouyeNAT98} or optical \cite{FatemiPRL00} fields. The FR
mechanism allows for a practical means to manipulate the nonlinearity
\cite{Malomed06}.

Although the control of the nonlinearity has been very effective,
noises can appear in the system management or added to it. In this
way, the perturbations can change the nonlinearity, and thus influencing
possible changes in the noise-free solutions. The inclusion of
spatial random potential in the study
of BEC dynamics has been proposed in Refs. \cite%
{LyePRL05,ChenPRA08,AbdullaevPRA05,AkkermansJPB08} and recently was
tested experimentally \cite{SchwartzNAT07}. When randomness is introduced
in the BEC dynamics, through an optical speckle, one may be able
to study Anderson localization in the context of BEC and superfluidity \cite{LyePRL05}.
It was clearly demonstrated in these papers that in the presence of disorder the condensate's 
expansion in 1D waveguides is inhibited, and the collective dipole and quadrupole oscillations
are strongly damped. In a way, this is similar to the damping of collective states in nuclei and metal clusters \cite{HussAP00}, where
the randomness is internal.

A natural question arises as to what would be the consequence of having the disorder present
directly in the nonlinearity term? For BEC, this implies a point two-body interaction (t-matrix) with a
random component. Would this add or remove some of the effects of the speckle potential?
In this connection, 
\cite{ChenPLA09}  have recently proposed an NLSE in the presence of random
nonlinearity. Specifically, these authors consider the effects of random time
modulation of the nonlinearity coefficient on the dynamics of
solitary waves in the NLSE. On the other hand, to our knowledge,
chaotic perturbations in the nonlinearity term in the NLSE have not been fully
considered yet, though it is a common knowledge that several physical systems
do exhibit chaotic behavior. 

Classically chaotic systems appear to  behave as random systems.  
Tiny differences in the initial state of the system can lead to
enormous differences in the final state even over
fairly small time scales \cite{Devaney89}. This happens even 
though these systems are deterministic, meaning that their future dynamics are fully
determined by their initial conditions with no random elements
involved. This behavior is known as deterministic chaos.

The dynamical systems theory (DST) is an area
whose interest lies mainly in nonlinear phenomena, the
source of classical chaos. DST groups use several
concepts to the study of chaos, such as Lyapunov
exponents, fractal dimension, bifurcation, and symbolic
dynamics among other elements \cite{Devaney89}. Recently,
other approaches have been considered, such as information
dynamics and entropic chaos degree \cite{OhyaIJTP98}. For example, given $C_0$, $C_n$ can be the $n$-th iterate
of the quadratic functions: $C_n(\mu) = C^2_{n-1} + \mu$; sine
functions: $C_n(\mu) = \mu \sin(C_n-1)$; logistic functions
$C_n(\mu) = \mu C_{n-1}(1-C_{n-1})$; exponential functions:
$C_n(\mu) = \mu \exp(C_{n-1})$; doubling function defined on the
interval $[0,1)$: $C_n = 2C_{n-1}~ mod~ 1$, and so on, $\mu$ being a
parameter. It is worth recalling that all the functions
in the above list are familiar to researchers in DST. For example, for
some values of $\mu$, it is known that some of these functions
can behave in quite a chaotic manner \cite{Devaney89}. In what follows
we make a distinction between chaos and randomness, though both
concepts indicate disorder.

The major thrust of our paper is to verify the influence of the
different types of perturbations in the nonlinearity of a system
governed by NLSE. In this sense, we investigate the chaotic, random,
and nonperiodic nonlinearity perturbation. We know that these
perturbations are different, however, to what extent the overall
effect is universal (independent on the details of the random perturbations), and  how can it modify a solitonic
solution? We purports to supply some answers to the above. 

Differently from Ref. \cite{ChenPLA09}, which considers a random
time modulation on a certain point (generating a Gaussian
distribution), here we consider a constant background nonlinearity
perturbed by a random function that interferes in both spatial and
temporal coordinates. Surprisingly, some of the solutions found here
can move depending of the amount of perturbed points in the
nonlinearity. This fact is similar to
those studied by the thermal effects on the nonlinearity \cite%
{GharaatiAPPA07}.

Firstly we consider the NLSE given by%
\begin{equation}
i\psi _{t}=-\psi _{xx}+g(x,t)\left\vert \psi \right\vert ^{2}\psi,  \label{NLSE}
\end{equation}%
where $\psi =\psi (x,t)$, $x$, and $t$ dimensionless, and $g(x,t)$ is the function
that describe the nonlinearity of the system. Here we consider
\begin{eqnarray}
g(x,t)&=&G(1+\sigma (x,t)),
\label{g}
\end{eqnarray}
where $G$ and $G\sigma (x,t)$ are the nonlinear parameter and the coefficient
generated by a chaotic, random, or nonperiodic generator, respectively. Eq.~(\ref{NLSE}) describes, 
e.g., the density of particles in a Bose-Einstein condensates when it is free of external 
potentials, in a configuration type cigar-shaped; the spatial pulse propagation in bulk crystals that 
present Kerr-effect in the nonlinearity or temporal pulse propagation in nonlinear optical fiber; etc.

We investigate the evolution of the solution for the NLSE
via numerical simulations, based on the split-step finite differences
(SSFD) method with a time-step and space-step size of $\Delta t=0.0001$ and $%
\Delta x=0.01$, respectively; we use $N$ to represent the number of points in space.
The core of SSFD is based on the Crank-Nicolson algorithm \cite{Vesely94}. 
To control these numerical simulations we looked for the conserved quantity
given by (power)
\begin{equation}
P=\sum_{x=1}^N\left\vert \psi(x,t) \right\vert ^{2}.
\end{equation}

To calculate the error we use the comparative form
\begin{equation}
E_{r}=\frac1N\sum_{x=1}^N\left( \left\vert \psi (x,0)\right\vert
^{2}-\left\vert \psi (x,t_{f})\right\vert ^{2}\right),
\end{equation}
where $t_f$ is the final time of the evolution. The equation above represents a mean distance between the input and 
output state.

When we take $G=-2$ and $\sigma(x,t)=0$ in (\ref{g}), we can write
\begin{equation}
\psi =e^{i\mu t}\mathrm{sech}(x)%
\label{in}
\end{equation}
as solution of (\ref{NLSE}), with $\mu=1$. This solution will be taken as
initial condition for our simulations of the Eq.~(\ref{NLSE}).

\begin{figure}
\includegraphics[width=2.8cm]{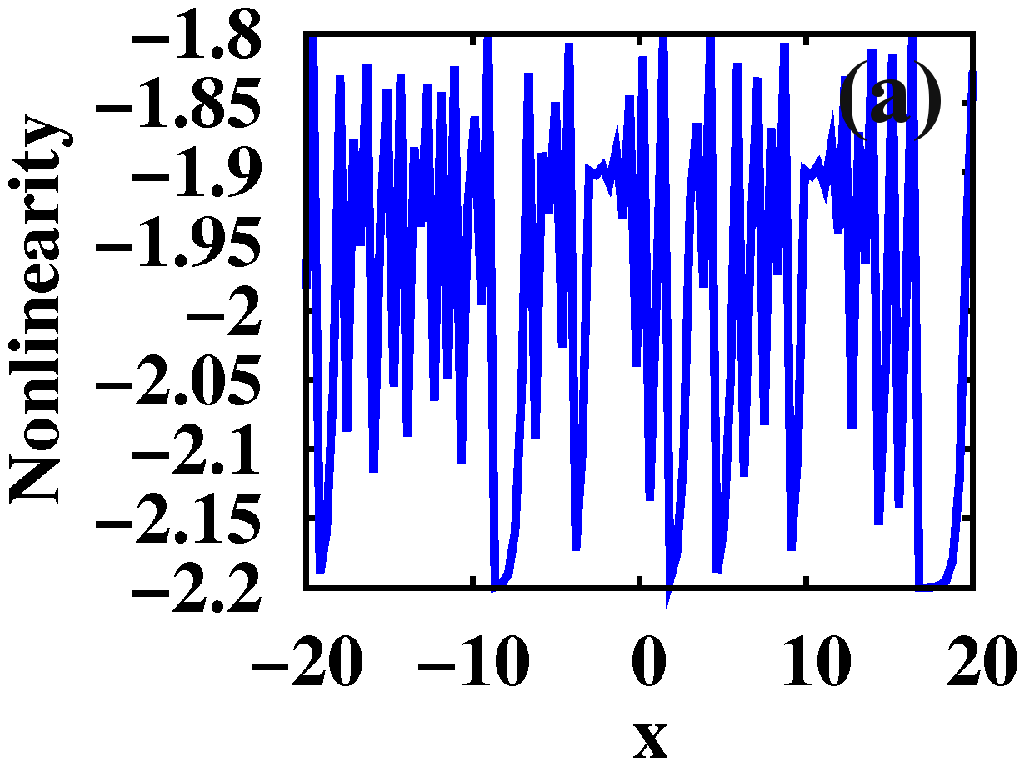} \hfil
\includegraphics[width=2.8cm]{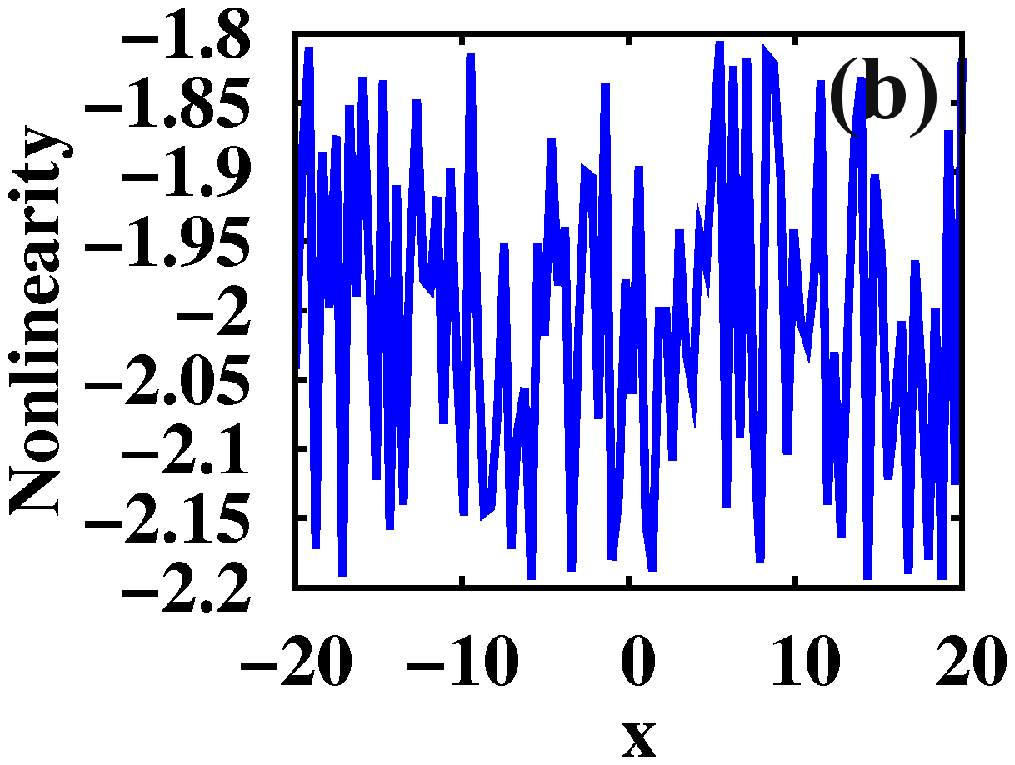} \hfil
\includegraphics[width=2.8cm]{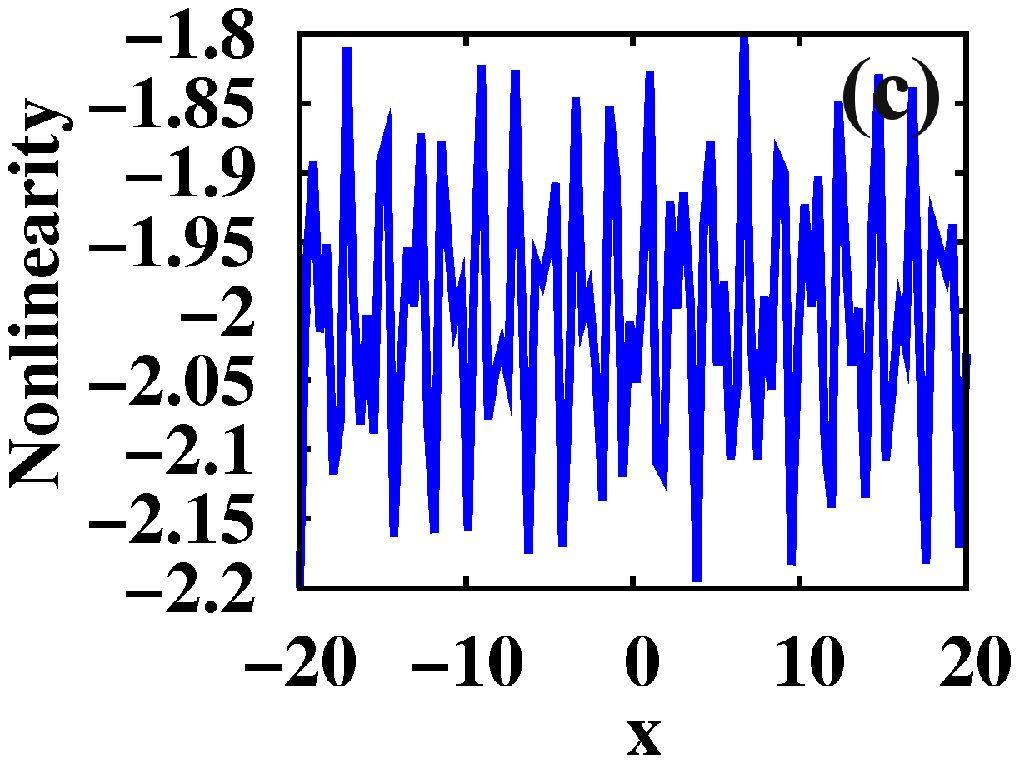}%
\caption{Plots of the perturbed nonlinearity for the (a) chaotic, (b) random, (c) and nonperiodic functions.}
\label{F1}
\end{figure}

For our numerical simulations we consider the chaotic perturbation in
the nonlinearity given by the logistic function, the random
perturbation is generated by random algorithm simulator, and the
nonperiodic perturbations are generated by the function
$\alpha(\cos(5x)/2+\cos(\sqrt{5}x)/2)$, where $\alpha$ assumes the
values of the perturbation. In Figs. \ref{F1}a, \ref{F1}b, and
\ref{F1}c we display a form of the chaotic, random, nonperiodic
nonlinearity as functions of space for a generic time, respectively.
Experimentally, this perturbation in the nonlinearity can be
constructed, e.g., in a crystal with impurities which are altered chaotically,
randomly, or non-periodically. It remains to be seen what the presence of
$G\sigma(x,t)$ implies, though one would guess that it amounts to taking into account, 
within the mean field, Gross-Pitaevskii, description, the effects of the many-body correlations.

\begin{figure}
\includegraphics[width=4.2cm]{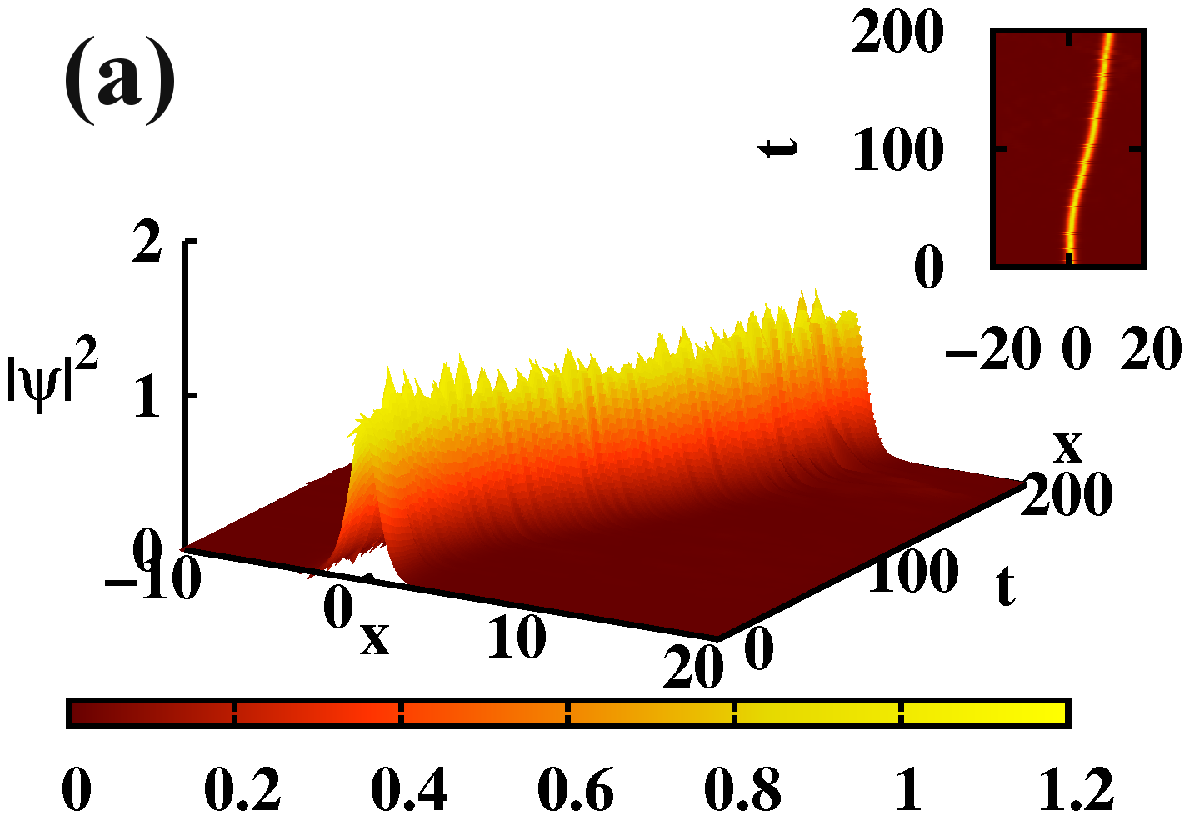} \hfil
\includegraphics[width=3.8cm]{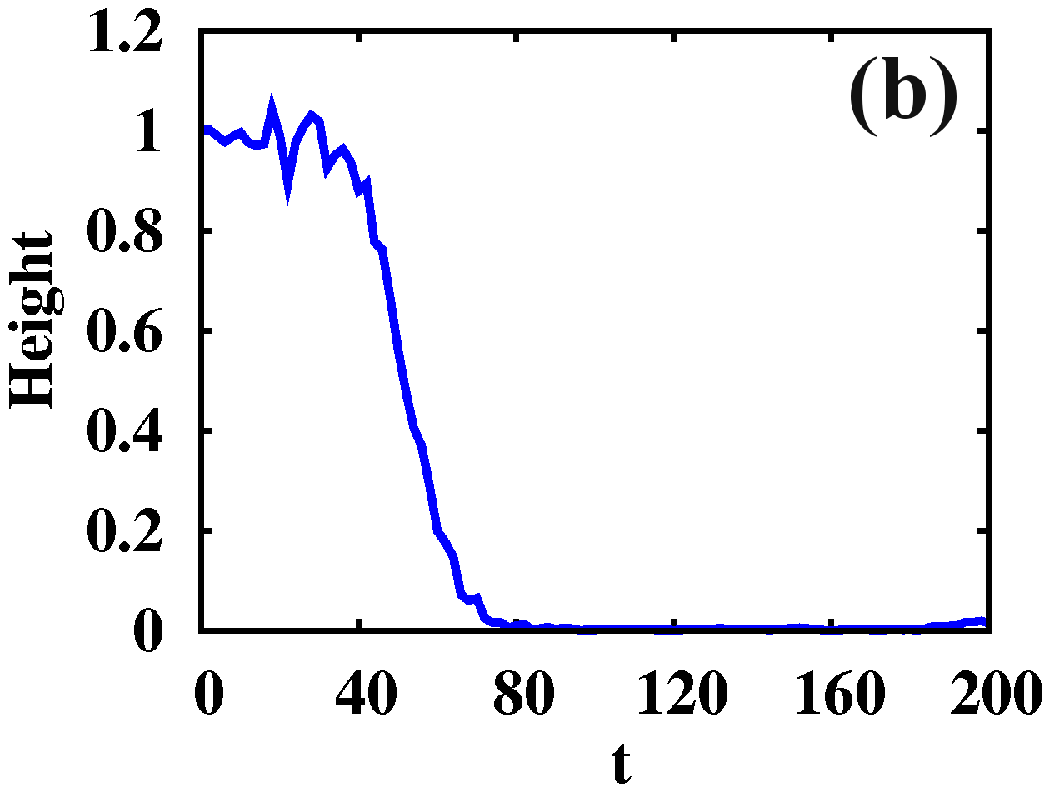} %
\caption{Plots of the soliton evolution $|\psi(x,t)|^2$ with $10\%$ of chaotic perturbation in the
nonlinearity. In (a) is displayed the solution and its profile (top panel) and (b) its height at position $x=0$.}
\label{F2}
\end{figure}

We use (\ref{in}) as input state in (\ref{NLSE}) to verify
its evolution in the presence of the above mentioned perturbations. In Fig.~\ref{F2}a we
consider the chaotic perturbation via $\sigma (x,t)$ between
$\pm0.001$ and $\pm0.1$ ($10\%$). The chaotic perturbation is
obtained considering $4000$ points affected within the interval
$-20\leq x\leq20$ in space, that changes the spatial profile of the
nonlinearity. These perturbed points are changed by a new function
after a time $t=20$, and so we will have $100$ temporal points
affected in the interval $0\leq t \leq200$. Fig.~\ref{F2}b displays
the soliton amplitude (height) of the solution ($|\psi|^2$) at the
position $x=0$. Note that it becomes vanishingly small after $t=60$ owing to
its motion. However, the soliton is stable in this range of
perturbation. We calculated the error in the power of
$2.41\times10^{-4}$ and the comparative error of $E_r=9.39$.

\begin{figure}
\includegraphics[width=4.2cm]{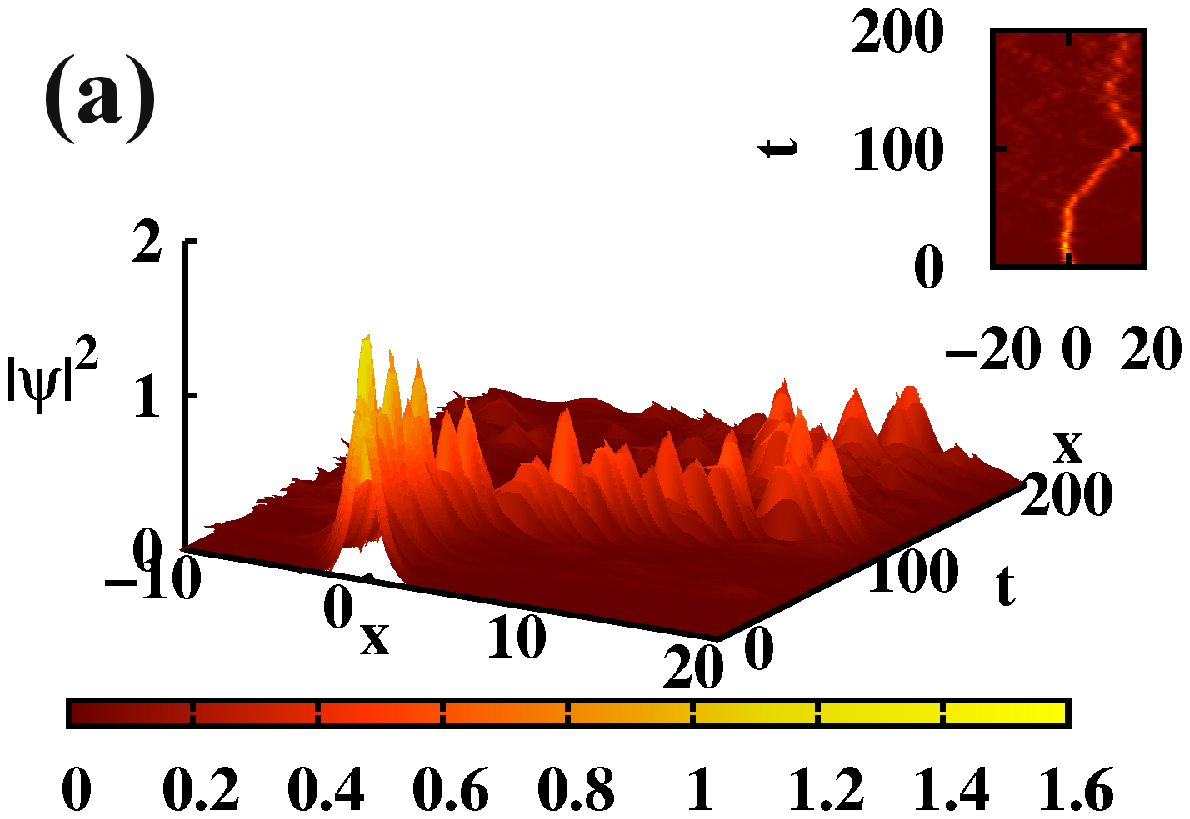} \hfil
\includegraphics[width=3.8cm]{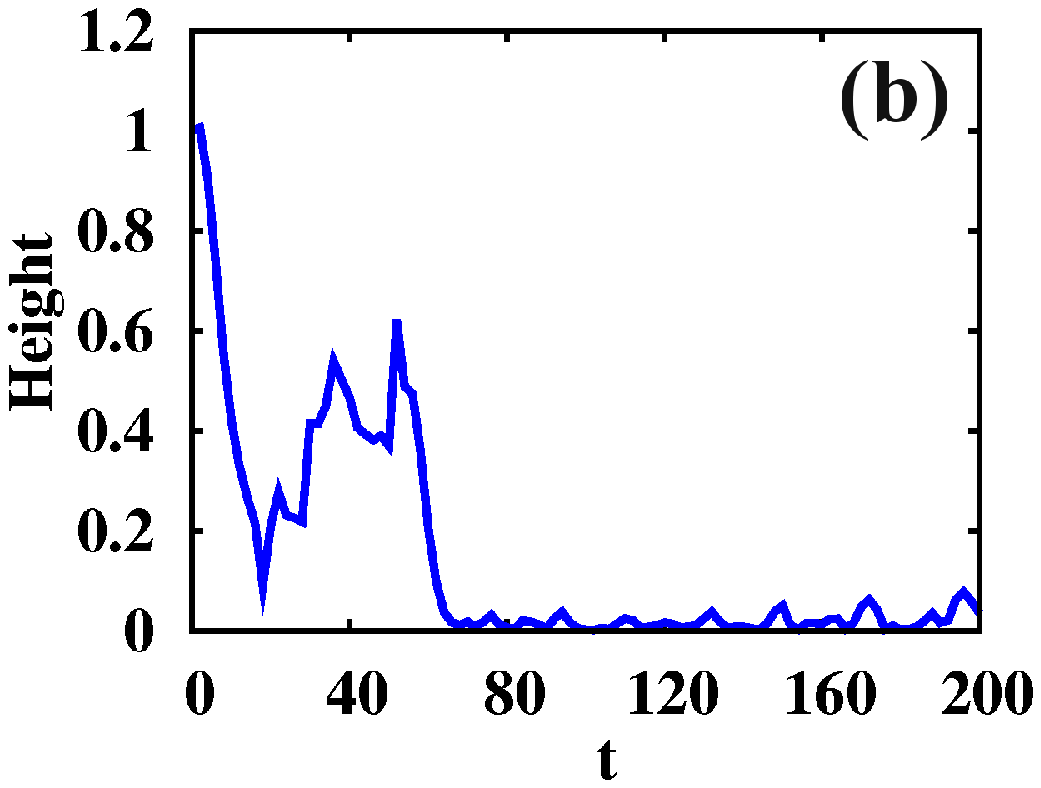} %
\caption{Plots of the soliton evolution $|\psi(x,t)|^2$ with $50\%$ of chaotic perturbation in the
nonlinearity (see text for details). In (a) is displayed the solution and its profile (top panel) and (b) its height at position $x=0$.}
\label{F3}
\end{figure}

On the other hand, when we consider a chaotic perturbation of $50\%$ in the value of the
amplitude, we found that the soliton practically disappears.
 This case is shown in Fig. \ref{F3}. In the Figs. \ref{F3}a
and \ref{F3}b we display the $|\psi|^2$ and the height in the position $x=0$%
, respectively. The error in the power was $2.38\times10^{-4}$ with a mean
distance $E_r=9.76$. In this case the mean distance is due to the moving
and the vanishing pattern, i.e., the output state can not be at the same position as that of the input
state.

\begin{figure}
\includegraphics[width=4.2cm]{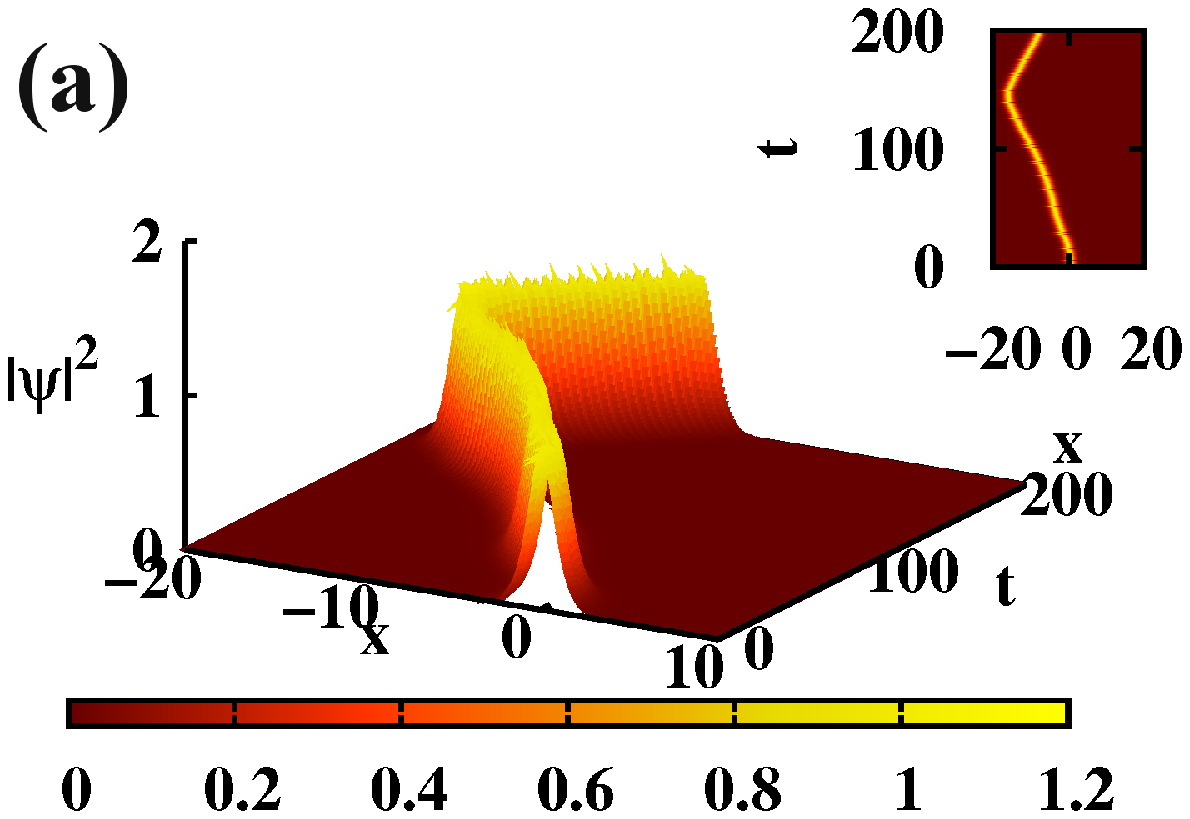} \hfil
\includegraphics[width=3.8cm]{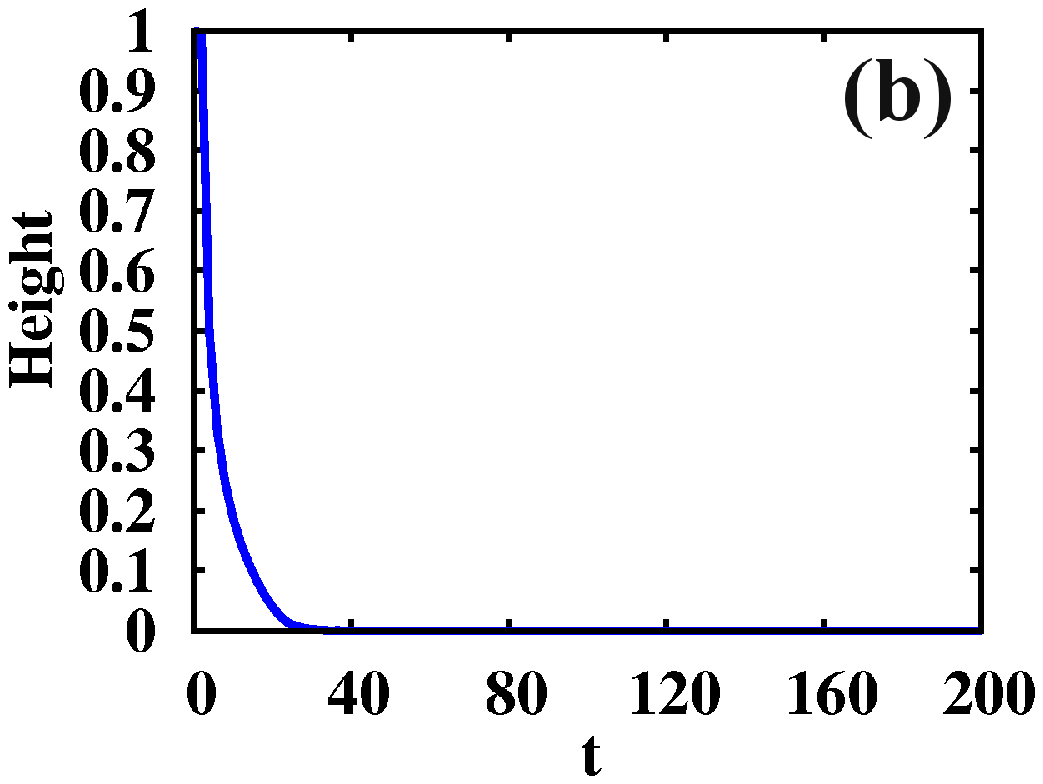} %
\caption{Plots of the soliton evolution $|\psi(x,t)|^2$ with $10\%$ of random perturbation in the
nonlinearity (see text for details). In (a) is displayed the solution and its profile (top panel) and (b) its height at position $x=0$.}
\label{F4}
\end{figure}

\begin{figure}
\includegraphics[width=4.2cm]{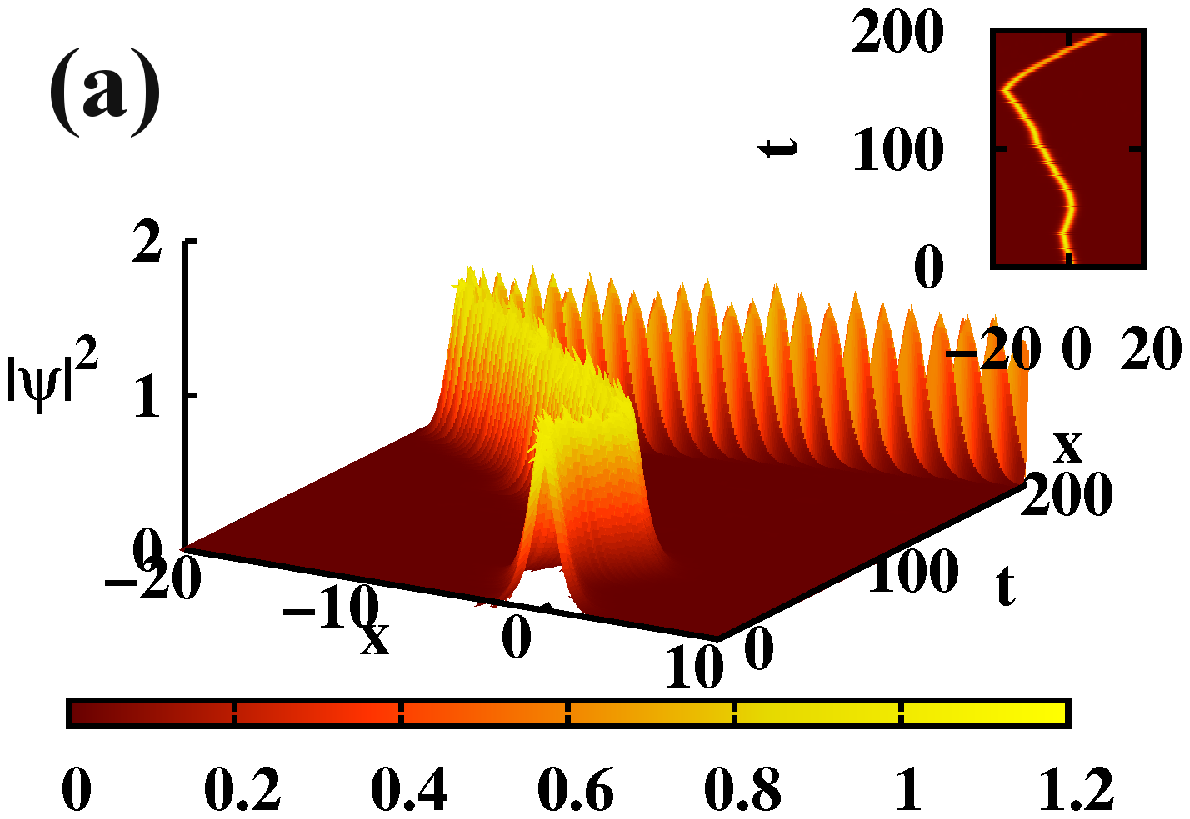} \hfil
\includegraphics[width=3.8cm]{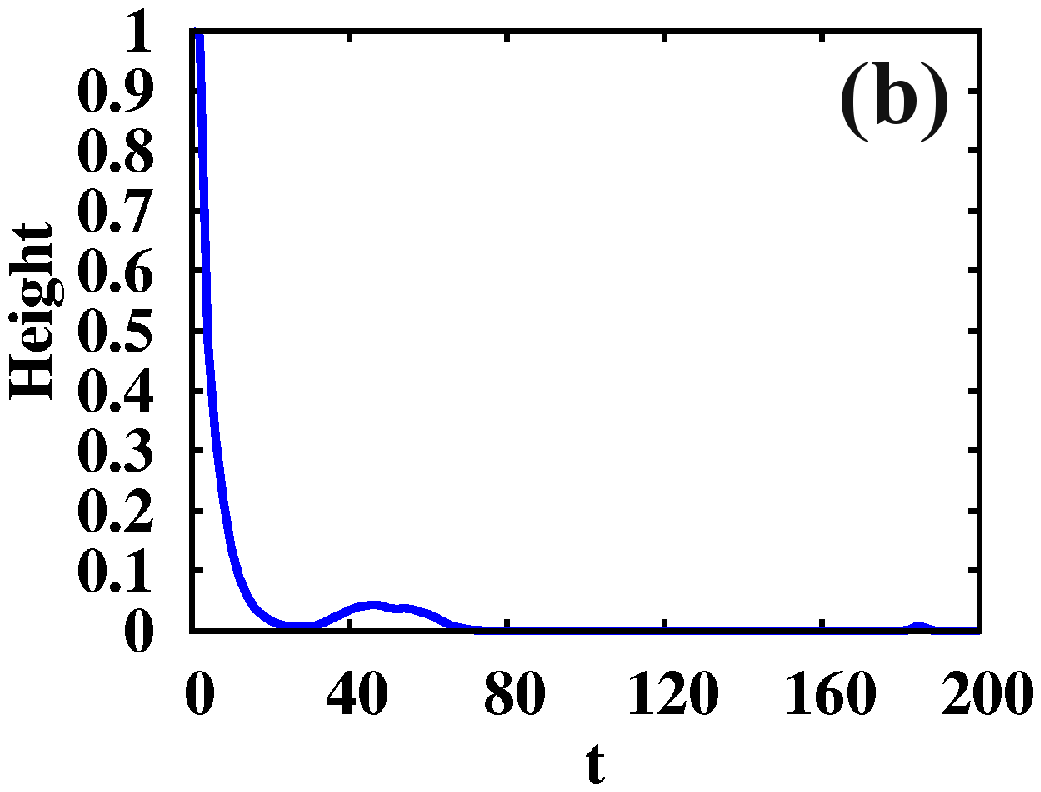} %
\caption{Plots of the soliton evolution $|\psi(x,t)|^2$ with $50\%$ of random perturbation in the
nonlinearity (see text for details). In (a) is displayed the solution and its profile (top panel) and (b) its height at position $x=0$.}
\label{F5}
\end{figure}

Now, when we consider a random perturbation of the nonlinearity,
obtained here by an algorithm of random number generation, we note a
different behavior of the solutions. Here the soliton remains stable
even for $50\%$ of perturbation, differently from the case of the
chaotic perturbation. In Fig.~\ref{F4}a we plot the soliton solution
considering $10\%$ of random perturbation in the nonlinearity using
the same arguments presented for the chaotic case, i.e., $4000$
affected points in space versus $100$ temporal points into the range
shown. Fig.~\ref{F4}b displays the amplitude of the soliton at $x=0$.
This perturbation is responsible for moving the soliton. The error in
the power is of $4.88\times10^{-6}$ with the mean distance between
the input and output states of $10^{-1}$. For $50\%$ of random
perturbation the oscillation of the soliton is more evidenced when
compared with the case of $10\%$ of random perturbation. Figs.~\ref{F5}a
and \ref{F5}b show the $|\psi|^2$ and the amplitude for
$x=0$, respectively. The robustness is guaranteed, differently form
the chaotic case. The errors in power and the mean distance are
$1.17\times10^{-5}$ and $10^{-1}$, respectively.

\begin{figure}
\includegraphics[width=4.2cm]{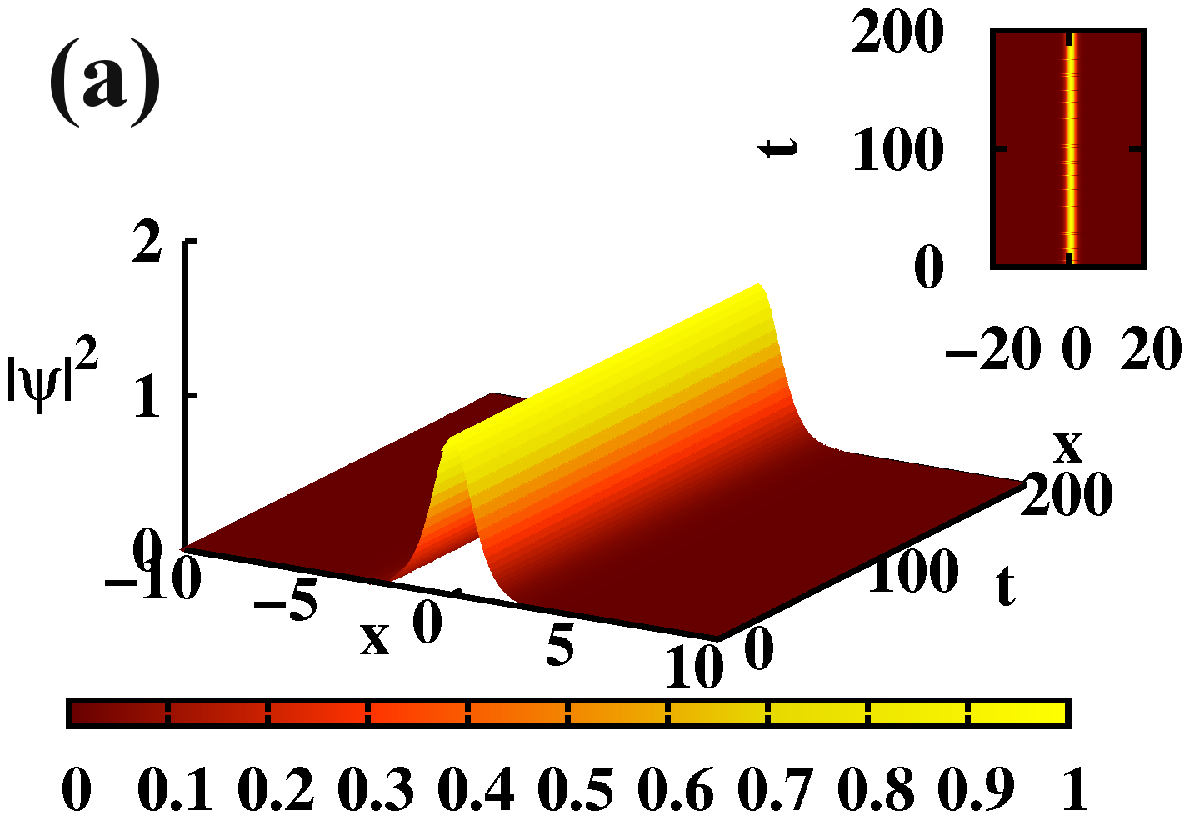} \hfil
\includegraphics[width=3.8cm]{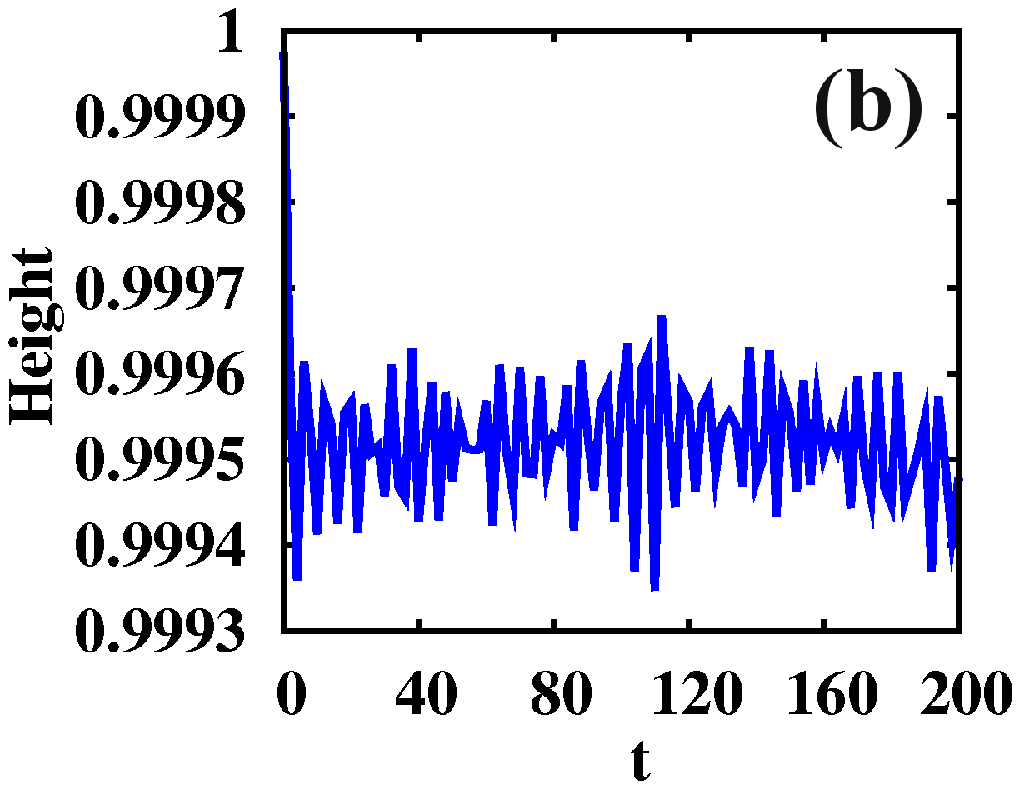} %
\caption{Plots of the soliton evolution $|\psi(x,t)|^2$ with $50\%$ of nonperiodic perturbation in the
nonlinearity (see text for details). In (a) is displayed the solution and its profile (top panel) and (b) its height at position $x=0$.}
\label{F6}
\end{figure}

To conclude our study we investigate the effects of the nonperiodic perturbation in
the nonlinearity. From Fig.~\ref{F1}c one observes that this
perturbation seems the most well-behaved compared to the other two types. This
fact is reflected in the behavior of the solution which remains practically with the
same form as that of the input state. This occurs even when it suffers
$50\%$ of nonperiodic perturbation, as can be seen in Fig.~\ref{F6}.
The error in the power is of $2.83\times10^{-4}$ and the mean
distance is of $2.88$.

With the results presented here, we verify that, under similar
conditions, chaotic, random, and nonperiodic perturbations in the
nonlinearity can present distinct features, and sometimes results in vanishing solitons,
as verified when the system suffers chaotic perturbations.

In summary, in the present work we have studied the effects of chaotic, random, and nonperiodic perturbations in the nonlinearity
on the soliton evolution via NLSE. We considered cubic nonlinearity with strength
perturbed chaotically, randomly, or nonperiodically. In the chaotic case we found that moving solitons
can be destroyed when they are perturbed with $50\%$ in the intensity of the nonlinearity. On the other hand,
when the system engenders random perturbation this does not occur. The soliton solution remains stable, however now it can move.
Finally, when we look for the nonperiodic perturbation we found that
it displays robust solutions with no apparent influence on the solitons. In a way, we can say that disorder in the non-linearity
may, or may not lead to Anderson-type localization, depending on the nature of the perturbation. Our results have direct
impact on work in optical lattices with impurities in the crystal, laser-generated randomness in the non-linearity, and many-body effects
in the dynamics of Bose-Einstein condensates and their collective excitations and transport. 

\subsection*{Acknowledgments}

The authors would like to thank CAPES, CNPQ, FUNAPE-GO, and FAPESP for partial financial support.

\end{document}